# Quantum-assisted cluster analysis

Florian Neukart[*1], David Von Dollen[1], Christian Seidel[2]

[1]Volkswagen Group, Region Americas
[2]Volkswagen Data:Lab

**Abstract**

We present an algorithm for quantum-assisted cluster analysis (QACA) that makes use of the topological properties of a D-Wave 2000Q quantum processing unit (QPU). Clustering is a form of unsupervised machine learning, where instances are organized into groups whose members share similarities. The assignments are, in contrast to classification, not known a priori, but generated by the algorithm. We explain how the problem can be expressed as a quadratic unconstrained binary optimization (QUBO) problem, and show that the introduced quantum-assisted clustering algorithm is, regarding accuracy, equivalent to commonly used classical clustering algorithms. Quantum annealing algorithms belong to the class of metaheuristic tools, applicable for solving binary optimization problems. Hardware implementations of quantum annealing, such as the quantum annealing machines produced by D-Wave Systems [1], have been subject to multiple analyses in research, with the aim of characterizing the technology's usefulness for optimization, sampling, and clustering [2–16, 38]. Our first and foremost aim is to explain how to represent and solve parts of these problems with the help of the QPU, and not to prove supremacy over every existing classical clustering algorithm.

## 1   Introduction

Quantum annealing is a class of algorithmic methods and metaheuristic tools for solving search or optimization problems. The search space for these problems usually consists of finding a minimum or maximum of a cost function. In searching a solution space for a problem, quantum annealing leverages quantum-mechanical superposition of states, where the system follows a time-dependent evolution, where the amplitudes of candidate states change in accordance of the strength of the transverse field, which allows for quantum tunneling between states. Following an adiabatic process, a Hamiltonian is found whose ground state closely describes a solution to the problem [1,2,28].

Quantum annealing machines produced by D-Wave Systems leverage quantum annealing via its quantum processor or QPU. The QPU is designed to solve an Ising model, which is equivalent to solving quadratic unconstrained binary optimization (QUBO) problems, where each qubit represents a variable, and couplers between qubits represent the costs associated with qubit pairs. The QPU is a physical implementation of an undirected graph with qubits as vertices and couplers as edges between them. The functional form of the QUBO that the QPU is designed to minimize is:

---

[*] Corresponding author: florian.neukart@vw.com

$$Obj(x, Q) = x^T \cdot Q \cdot x$$

(1)

where $x$ is a vector of binary variables of size $N$, and $Q$ is an $N \times N$ real-valued matrix describing the relationship between the variables. Given the matrix $Q$, finding binary variable assignments to minimize the objective function in Equation 2 is equivalent to minimizing an Ising model, a known NP-hard problem [16,17].

## 2 Classical clustering

In cluster analysis, the aim is to group sets of objects, i.e., points or vectors in $d$-dimensional space, such that some objects within one group can be clearly distinguished from objects in another group. An additional task may be the ability to quickly assign new objects to existing groups (clusters), i.e., by calculating the distance to a previously calculated cluster-centroid instead of running the re-running the complete clustering algorithm.

Clustering is a form of unsupervised machine learning, and used to find representative cases within a data set for supporting data reduction, or when needing to identify data not belonging to any of the found clusters [29]. Clustering helps to identify instances similar to one another, and to assign similar instances to a candidate cluster. A set of clusters is considered to be of high quality if the similarity between clusters is low, yet the similarity of instances within a cluster is high [30]. The groups are, in contrary to classification, not known a priori, but produced by the respective clustering algorithm [31]. Clustering is, amongst others, supported by self-organizing feature maps, centroid-based algorithms [32], distribution-based algorithms, density-based algorithms, orthogonal partitioning clustering.

We only explain one very common algorithm in detail – self-organizing feature maps – as this classical algorithm shares some similarities to the introduced quantum-assisted clustering algorithm.

### 2.1 Self-organizing feature map

Self-organizing feature maps (SOFMs) are used to project high-dimensional data onto a low-dimensional map while trying preserve the neighboring structure of data. This means that data close in distance in an $n$-dimensional space should also stay close in distance in the low-dimensional map – the neighboring structure is kept. SOFMs inventor, Teuvo Kohonen, was inspired by the sensory and motor parts of the human brain [33].

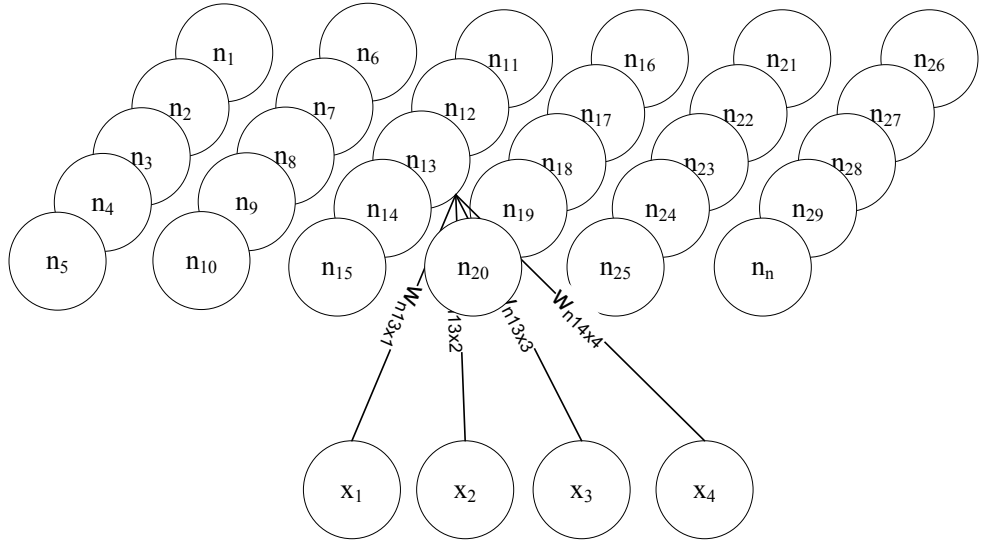

Fig. 1 – Self organizing feature map

Fig. 1 – Self organizing feature map – the scheme of a SOFM shows that every component of the input vector $x$ is represented by an input neuron and is connected with the above low-(two-) dimensional layer. During a learning phase, the weight vectors of a SOFM are adapted in a self-organizing way [34]. As other Artificial Neural Networks (ANNs), the SOFM consists of neurons ($n_1, \dots, n_n$), each having a weight vector $w_i$ and a distance to a neighbor neuron. The distance between the neurons $n_i$ and $n_j$ is $n_{ij}$. As Fig. 1 – shows, each neuron is allocated a position in the low-dimensional map space. As in all other ANNs, initially the neuron weights are randomized. During learning, the similarity of each input vector to the weights of all neurons on the map is calculated, meaning that all weight vectors are compared with the input vector $d \in D$. The SOMs learning algorithm therefore belongs to the group of unsupervised learning algorithms. The neuron showing the highest similarity, having the smallest distance $d_{small}$ to $d \in D$ is then selected as the winning neuron $n_{win}$ (Eq. 3) [35]:

$$d_{small} = {\min_{1 \leq j \leq n}} d\{(d \in D, w_j)\}$$

(2)

Weights of the winning neuron are adapted, as well as the weights of the neighbor neurons utilizing the neighborhood function $\varphi_n$ and the learning rate $\mu$. The neighborhood function has the following characteristics [35]:

- $\mu$ has its center at the position of $n_{win}$ and is a maximum there.
- The neighboring neurons are considered according to a radius. Within this radius, for distances smaller than $r$, $\varphi_n$ leads to outcomes greater than zero, and for distances greater than $r$, it takes on a value of zero.

Choosing a Gaussian function fulfils all the requirements in this case. The adaption of the weights is then carried out as described in Eq. 3:

$$w_i^{(t+1)} = w_i^{(t)} + \mu \varphi_n\left(w_{n_{win}}, w_i^{(t)}, r\right)\left(d \in D - w_i^{(t)}\right)$$

(3)

During training, the learning rate and the neighborhood radius has to be reduced in each iteration, done by $\sigma^{(t+1)}$ (Eq. 4) [35, 37]:

$$\sigma^{(t+1)} = \sigma_s * \left(\frac{\sigma_e}{\sigma_s}\right)^{(t+1)/(t+1)_e}$$

(4)

where $\sigma_s$ represents the starting value and $\sigma_e$ the ending value, also being the function value of $t(+1)_e$.

## 2.2 Similarities to SOFM and quantum-assisted clustering

In the example depicted in Fig. 1, the SOFM is a two-dimensional lattice of nodes, and depending on a presented instance, different nodes will fire with different strengths. The ones firing with the greatest amplitude give the cluster assignment. The QACA works similar in the sense that the two-dimensional topological properties of the D-Wave are exploited for cluster assignments. Assuming we embed two-dimensional clusters on the chip (higher-dimensional structures can be mapped as well – see the explanations in chapter 3), an assignment of cluster points to qubits may look as described in Fig. 2:

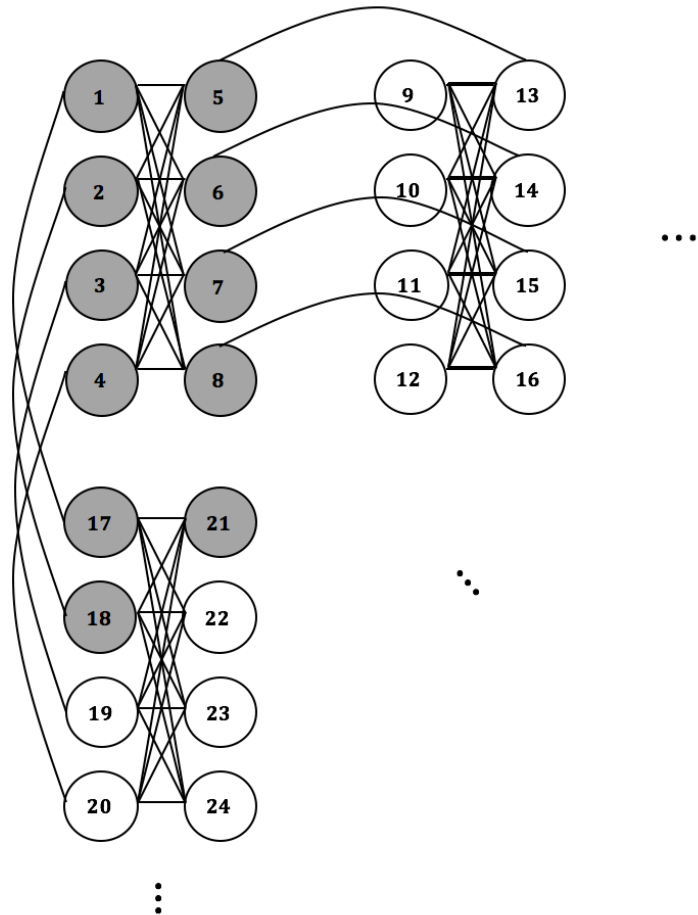

Fig. 2 – Qubits and clusters

Fig. 2 shows schematically that qubits 1 – 8, and 17, 18, 21 would "fire", thus take the value 1 in the result-vector, and qubits 9 – 16 and 19, 20, 22 – 24 would not fire, thus take the value 0. We need to set the couplings accordingly, so that when a candidate instance is fed into the cluster-form (see 3.2, Fig. 3) and embedded onto the QPU, the result allows us identify "areas" of activity or groups of qubits set to 1 for similar instances.

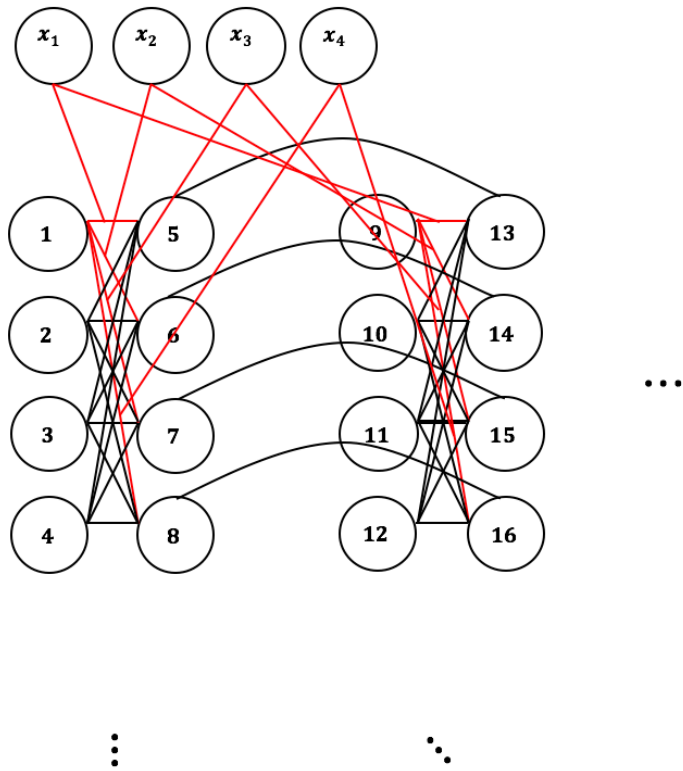

Fig. 3 – Feeding an instance into the cluster-form

Fig. 3 shows how an instance is fed into the cluster-form. $\vec{X} = (x_1, \ldots, x_n)$ represents the input vector.

## 3 Quantum-assisted clustering analysis (QACA)

The introduced algorithm can be used as a probabilistic and definite clustering-algorithm, depending on how the result-vector is interpreted.

### 3.1 Quantum-assisted clustering with $n$-dimensional polytypes

The underlying idea is to classically define $n$-dimensional polytypes, such as the tetrahedron, the pentachoron, the tesseract, or even typeless polygons, which serve as clusters into which the instances projected, and map these onto the two-dimensional graph of the quantum annealing chip. The structure is derived from the number of input attributes in the data set. If each instance comes with 3 attributes, the structure of choice is a tetrahedron, and if the number of input attributes is 4, the structure of choice is a tesseract (see Fig. 4).

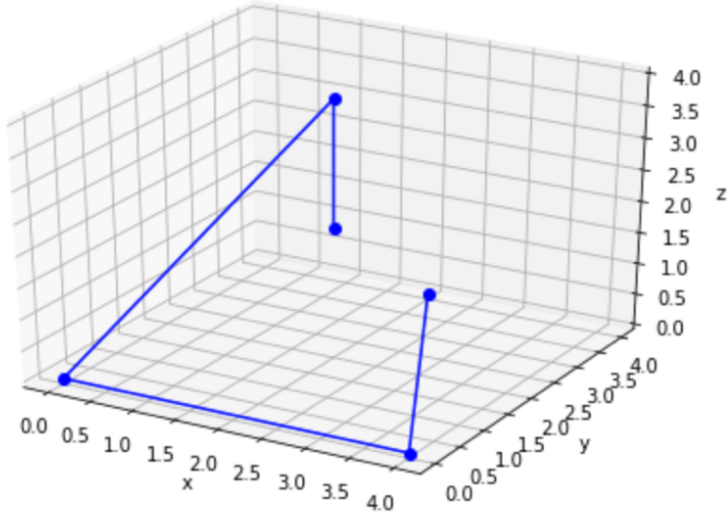

Fig. 4: Tetrahedron in three dimensions

The tetrahedron in three dimensions is given by four vertices, and assuming the intention is to cluster a four-dimensional data set into three clusters, three tetrahedra need to be defined. We do this by generating three random centroids, from which we calculate the remaining vertices. The centroid of the tetrahedron in Fig. 4 is given by the coordinates $c = (2,2,2)$. The remaining coordinates can be easily calculated, depending on the desired cluster size. Assuming we define a distance of 2 from the centroid, the set of tetrahedral coordinates $P = \{p_1, p_2, p_3, p_4\}$ are calculated as described in Eqs. 5 – 9:

$$p_1 = (c_x, c_y, c_z + 2)$$

(5)

$$p_2 = (c_x - 2, c_y - 2, c_z - 2)$$

(6)

$$p_3 = (c_x + 2, c_y - 2, c_z - 2)$$

(7)

$$p_4 = (c_x, c_y + 2, c_z - 2)$$

(8)

where the centroid $c$ is defined as

$$c = (c_x, c_y, c_z)$$

As this approach does not generalize to other polytypes, the three-dimensional tetrahedron serves only as an example. Another way of defining clusters is by typeless polygons, based on randomly chosen coordinates from within a range of $min(x)$ and $max(x)$. Due to the inner workings of the introduced algorithm strongly overlapping clusters can be seen as probabilistic clustering, and clusters within clusters would help to identify clusters in data sets such as described in Fig. 5:

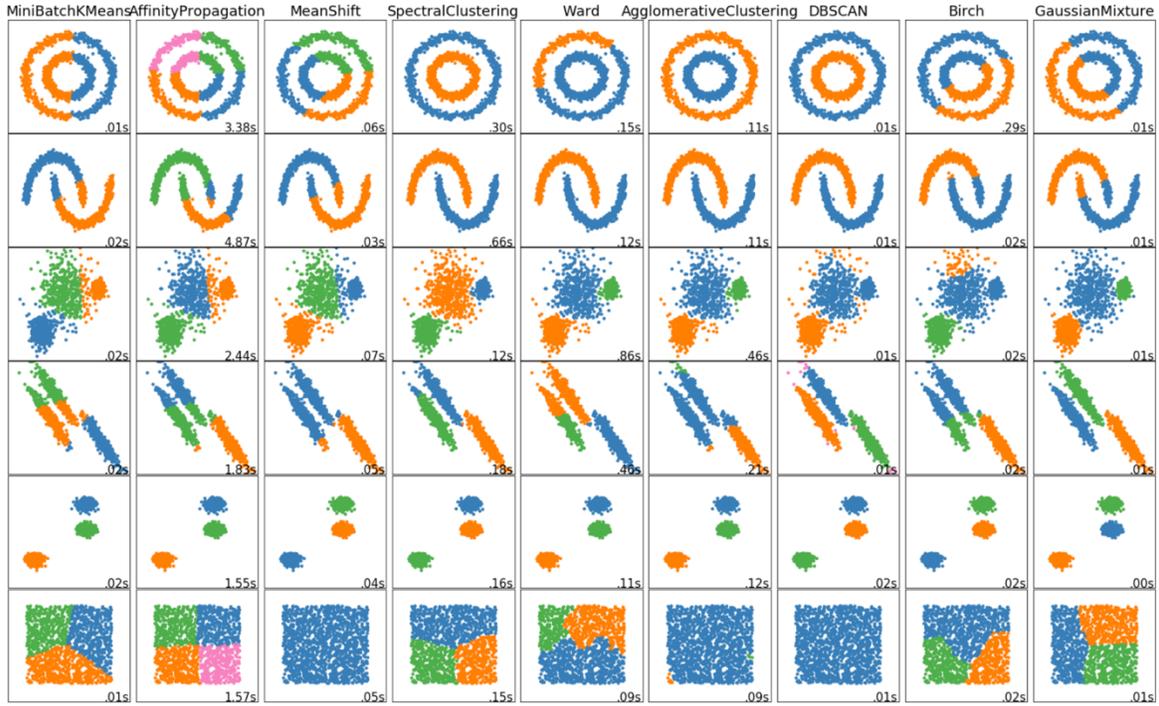

Fig. 5: Some non-linear data sets and some ways to cluster them [36]

Depending on how far we move the clusters apart, the less probabilistic QACA becomes, as the farther the clusters are apart, the smaller the probability of overlapping clusters becomes. If, classically (non-quantum), clusters do not overlap at all, we find definite cluster assignments for each of the instances. To give a first indication about how we define probability in terms of the introduced quantum-assisted clustering algorithm, we consider definite states of qubits post-measurement. Each qubit can be in one of the states $S = \{-1,1\}$. The more qubits of a cluster $k_x \in K = \{k_0, \dots, k_{m-1}\}$ take the state 1 for a specific instance $i_x \in I = \{i_0, \dots, i_{l-1}\}$, the more probable it is that the instance $i_x$ is a member of $k_x$. What's particularly elegant about this approach is that if clusters do not overlap in space, the nature of our algorithm still allows for probabilistic clustering (and to solve non-linear problems as depicted in Fig. 3). However, the farther apart we move the clusters, the more the respective cluster coordinates differ from each other, and the more likely it is that we find definite assignments. We initialize the clusters based on $n$-dimensional typeless polygons as described in Algorithm 1:

**Algorithm 1** Cluster definition based on $n$-dimensional typeless polygons

---

**Initialize:** $i_c, n_v, M, i_+, r_{min}, r_{max}$
**For each** $k \in M$:
    **For each** $v \in N_v$:
        $v_x^c = rand(r_{min}, r_{max})$
        $v_y^c = rand(r_{min}, r_{max})$
        $v_z^c = rand(r_{min}, r_{max})$
    $r_{min} = r_{min} + i_+ * \epsilon$
    $r_{max} = r_{max} + i_+ * \epsilon$

---

**Breakdown**
$i_c$: the initial coordinate for cluster vertex calculations, given by Eq. 8.
$n_v$: set of all vertices per cluster, i.e., four vertices per cluster: $N_v = \{1,2,3,4\}$.
$k$: cluster
$M$: set of all clusters, i.e., three clusters: $M = \{1,2,3\}$.
$i_+$: increment by which the coordinate range for finding random vertices is shifted, given by Eq. 9.
$r_{min}$: minimum range value for finding random vertices which define a cluster. Initialized as $r_{min} = i_c$.
$r_{max}$: maximum range value for finding random vertices which define a cluster. Initialized as $r_{max} = r_{min} + i_+$.
$v_x^c, v_y^c, v_z^c$: x, y, z coordinates of the vertex $v$ in the $c^{th}$ cluster. In the introduced example space is 3-dimensional, but the algorithm generalizes to $n$-dimensional space, and even complex manifolds.
$\epsilon$: sliding factor.

$$i_c = \min(X) \tag{10}$$

$$i_+ = \frac{\max(X) - \min(X)}{m} \tag{11}$$

where $X$ is the matrix of input attributes and $m$ the number of clusters. In Alg. 1, we assign coordinates to each vertex of an $n$-dimensional typeless polygon. For each cluster, we shift the coordinate range $r = (r_{min}, r_{max})$ by the increment $i_+$ and a sliding factor $\epsilon$, which is increases or decreases in coordination with desired inter-cluster distances. We emphasize that large inter-cluster distances, i.e. in the Euclidean sense, do not necessarily imply definite cluster assignments. For an instance $i_x$, the introduced algorithm may still calculate a certain probability of $i_x$ belonging to cluster $k_1$, but also to $k_x$, even when $k_1$ and $k_x$ do not overlap in $n$-dimensional space.

## 3.2 QUBO-form and embedding

We present the problem to the D-Wave in QUBO-form. The definition of the matrix in QUBO-form is done in two steps.

1. The first step is in defining a matrix in QUBO-form or what we call a cluster-form (CF). The CF is defined only once for all presented instances, and subsequently modified as instances are fed into it. It is worth pointing out another major difference to classical clustering algorithms such as k-means or self-organizing feature maps: instead of training regimes, i.e., iterative distance-based calculation of centroids, or strengthening the weights of nearest neighbors around a firing neuron, we only need to allocate instances to the CF once to obtain the cluster assignment.

   The QUBO-matrix is an upper triangular $N \times N$-matrix defined by $i \in \{0, \ldots, N-1\}$ by $j \in \{0, \ldots, N-1\}$. In the demonstrated example, each entry is initialized with 0, and subsequently updated with the values calculated for the CF, which come from Alg. 1. The CF will hold all values of the vertices based on the simple calculations in Alg. 1. While calculating each vertex coordinate $v_x^c, v_y^c, v_z^c$, we also assign an ID to each of these and store this information in a lookup-table. The $x$-coordinate in first vertex in the first cluster is given the ID 1: $v_x^1$ (or more accurately: $v_{1_x}^1$, where the exponent defines the cluster, and the subscript the vertex number and the respective coordinate), the $y$-coordinate in the first vertex of the first cluster the ID 2, and so on. We additionally create a list $L$ of length $l = n_v * m$, which contains a list of the coordinate values, i.e., the first three entries of this list give the $x, y, z$ coordinates of the first vertex in the first cluster. The values in $L$ may also be scaled as described in Eq. 20, but this strongly depends from the variance in the data set. We define the number of vertices as $n_v$ and $m$ the number of clusters. Additionally, we store the qubit-to-cluster assignments in a lookup-table $D$ in the form $\{k_1: [0,1,2], k_2: [3,4,5], \ldots, k_n: [q_{x-3}, \ldots, q_{x-1}]\}$ that we use in step 2. We assign $k_x$ as the cluster number, and qubits are given by the respective arrays. The CF is then defined as described in Eq. 12:

$$CF(i,j) = \begin{cases} CF(i,j) - \sqrt{(L_i^2 + L_j^2)}, & \text{if c1} \\ CF(i,j) + \sqrt{(L_i^2 + L_j^2)}, & \text{if c2} \\ CF(i,j), & \text{otherwise} \end{cases}$$

(12)

where

$$c1: S_1 \equiv S_2 \text{ and } i \leq j$$

(13)

and

$$c2: S_1 \neg\equiv S_2 \text{ and } i \leq j$$

(14)

In Eqs. 13 and 14 the conditions for assigning positive or negative signs to an entry are defined. If c1 is met, our tests show that setting the respective entries to 0 instead of $-\sqrt{(L_i^2 + L_j^2)}$ may provide better results, but there is a noticeable variance over differing data sets. The basic idea is to iterate over the qubit-IDs of each cluster, and to compare if the set of qubit IDs $S_1$ is identical to the set of qubit IDs $S_2$. If the sets are identical, negative intra-cluster couplings are set, and if not, positive inter-cluster couplings are set. The reason for this is that once we introduce an instance to the CF. The coupling-strengths values around the most probable cluster's qubits are lowered, and in the same instance the values the inter-cluster couplings help to raise the entries of the remaining clusters. This results in lower probability of the most probable clusters being activated.

2. The second step is iterating over all cluster-instances: the instances are fed into the cluster-form one by one, and each of the resulting instance-cluster matrices (ICM) are embedded on the QPU. For each cluster, we go over the number of vertices and calculate a distance from each attribute-coordinate to each cluster-coordinate. The number of qubits per cluster must be a multiple of the number of data set attributes, i.e., when the data set is three-dimensional, a cluster may be represented by 3 qubits (point), 6 qubits (line), 9 qubits(triangle), and so on. If a cluster in a 3-dimensional space is defined by 6 points, we require 18 qubits to represent it on the QPU. For each of the cluster coordinates, we now calculate the distance to each instance and update the list $L$ accordingly. $L$, as defined in step 1, was used to define the cluster-form and was set with negative intra-cluster couplings, and positive inter-cluster couplings. For each instance, $L$ is updated as described in Alg. 2:

**Algorithm 2** Instance to cluster distance calculation

**Load:** $D$, $L$, $i_x$
**Initialize:** $cc=0$
**For each** $k \in D$:
    **For each** $qubit \in k$:
        $L[qubit] = L[qubit] - i[cc]^2$
        $cc = cc + 1$
        **If** $cc == d$:
            $cc=0$

**Breakdown**
$D$: Cluster dictionary $D: \{k_1: [0,1,2], k_2: [3,4,5], \dots, k_n: [q_{x-3}, \dots, q_{x-1}]\}$
$L$: List with qubit-IDs and their values as initialized in the cluster-form
$i_x$: an instance
$cc$: coordinate counter. Counts up to 3 if the instance has 3 coordinates, up to 4 with 4 coordinates, and so on
$d$: number of dimensions per instance
$k$: key/ cluster in $D$
$qubit$: the qubit IDs per entry in $D$
$L[qubit]$: the value of $L$ at entry $qubit$
With Alg. 2, the distance from an instance $i_x$ to any point in any cluster in the cluster-form is calculated. Once this is done, the ICM is updated as described in Eqs. 15 – 20:

$$CF(i,j) = \begin{cases} CF(i,j) - (L_i^2 + L_j^2), & if\, c1 \\ CF(i,j) - (L_i * L_j), & if\, c2 \\ CF(i,j) + (L_i^2 + L_j^2), & if\, c3 \\ CF(i,j) + (L_i * L_j), & if\, c4 \\ CF(i,j), & otherwise \end{cases}$$

(15)

where

$$c1: S_1 \equiv S_2 \text{ and } i < j$$

(16)

and

$$c2: S_1 \equiv S_2 \text{ and } i = j$$

(17)

and

$$c3: S_1 \neg\equiv S_2 \text{ and } i < j$$

(18)

and

$$c3: S_1 \neg\equiv S_2 \text{ and } i = j$$

(19)

The last step before embedding the problem onto the QPU is scaling the values in the ICM, which is done according to Eq. 20:

$$x_{scaled} = \frac{x_i - mean(x)}{\sigma(x)}$$

(20)

where $\sigma(x)$ is the standard deviation. The features are centered to the mean and scaled to unit variance.

Once the ICM has been processed, the spin-directions provided in the result-vector tell us which qubits are "turned on", and which are "turned off". Three ways to extract the cluster assignments are probabilistic and definite:

1. Definite: For the turned-on qubits, the respective values of $L$ are extracted, and by looking up $D$ we can identify the cluster this qubit belongs to. In $D$, we can find the qubits per cluster, and from the result-vector we get the turned-on. We look up the respective IDs in $L$, and sum the values over the remaining qubits. The lowest sum of "on"-qubit values per cluster gives the cluster assignment.
2. Probabilistic 1: The number of turned-on qubits per cluster, as defined by qubit-assignments in $D$, is counted. The percentage of turned-on qubits per cluster gives the probabilistic assignments of an instance to clusters.
3. Probabilistic 2: For the turned-on qubits, the respective values of $L$ are extracted, and by looking up $D$ we can identify the cluster this qubit belongs to. In $D$, we can find the qubits per cluster, and from the result-vector we get the turned-on. We look up the respective IDs in $L$, and sum the values over the remaining qubits. The percentage of turned-on qubits-values per cluster gives the probabilistic assignments of an instance to clusters.

## 4  Experimental results and conclusions

Our intention was to obtain the results without having to split the QUBO so that a singular embedding is possible. We verified QACA with commonly used low-dimensional verification data sets, such as the Iris data set. For verification, we chose Expectation Maximization, k-means, and Self-Organizing Feature Maps, all three known to perform well on the Iris data set. We ran QACA 5 times and averaged the performance, as due to the randomness in the cluster-form the results can vary. In brackets, we provide the individual cluster assignments. The accuracy is defined as percentage of correctly assigned instances, and the cluster-assignment is definite (Tbl. 1).

|  | EM | k-means | SOFM | QACA |
|---|---|---|---|---|
| **Accuracy in %** | 86 | 89.7 | 70.7 | Avg.: ~85.6 <br><br> Ind.: (87.33 (131), 90 (135), 83.33 (125), 80 (120), 87.33 (131)) |

Tbl. 1 - Algorithm comparison

Some example results for the "Probabilistic 2"-method, which is as accurate as the definite results described in Tbl. 1 when assigning highest probability to an instance, are as follows (Tbl. 2):

```
instance 0 probabilities: 1.06, 20.96, 77.97
instance 1 probabilities: 1.06, 20.96, 77.97
instance 2 probabilities: 2.62, 20.99, 76.38
instance 3 probabilities: 0.76, 20.92, 79.83
instance 4 probabilities: 1.06, 20.96, 77.97
instance 5 probabilities: 4.019, 23.99, 80.02
...
```
Tbl. 2 – Probabilistic assignments

Summing up, the quantum-assisted clustering algorithm can compete with classical algorithms in terms of accuracy, and sometimes outperforms the ones used for comparison on the test data sets. However, the results strongly vary depending on the cluster-form, and better ways for cluster-form initialization have to be found.

# 5 Future work

In our future work, we intend to further exploit the chip topology to identify cluster assignments. By identifying where on the QPU we can find the turned-on qubits, an implementation of full feature map should be possible.

# Acknowledgments

Thanks go to VW Group CIO Martin Hofmann and VW Group Region Americas CIO Abdallah Shanti, who enable our research. Special thanks go to Sheir Yarkoni of D-Wave systems whose valuable feedback helped us to present our results comprehensibly.